\documentstyle[psfig]{l-aa}
\begin{document}
%%modificare \thesaurus{08(08.14.1; 08.16.7; 13.25.5)}
\title{Two--sided radio emission in ON 231 (W Comae)}
\author{E.~Massaro\inst{1}, F.~Mantovani\inst{2}, R.~Fanti\inst{2,3},
R.~Nesci\inst{1}, G.~Tosti\inst{4}, T.~Venturi\inst{2},
}
\institute{ Dipartimento di Fisica, Universit\`a "La Sapienza",
P.zle A. Moro 2, I-00185 Roma, Italy \and
Istituto di Radioastronomia, CNR, Via P. Gobetti 101, I-40129 Bologna,
Italy \and
Dipartimento di Fisica, Universit\`a di Bologna, Via Irnerio 46,
I-40126 Bologna \and
Osservatorio Astronomico, Universit\`a di Perugia, Via A. Pascoli,
I-06100
Perugia, Italy
}
\offprints{E.~Massaro: massaro@astro.uniroma1.it}
\date{Received 13 April 2000; accepted 17 May 2001}
\maketitle

\markboth{E. Massaro et al.: Two--sided radio emission in ON~231 }
{E. Massaro et al.: Two--sided radio emission in ON~231}

\begin{abstract}
Recent radio images of the BL Lac object ON~231 (W Com, 1219+285) show
remarkable new features in the source structure compared to
those previously published. The images were obtained from
observations made with the European VLBI Network plus MERLIN at
1.6 GHz and 5 GHz after the exceptional optical outburst occurred
in Spring 1998. The up-to-date B band historic light curve of
ON~231 is also presented together with the R band luminosity
evolution in the period 1994--1999. We identify the source core in the
radio images with the brightest component having the flattest spectrum.

A consequence of this assumption is the existence of a two--sided
emission in ON~231 not detected in previous VLBI images.
A further new feature is a large bend in the
jet at about 10 mas from the core. The emission extends for about
20 mas after the bend, which might be due to strong interaction
with the environment surrounding the nucleus. We suggest some
possible interpretations to relate the changes in the source
structure with the optical and radio flux density variation in
the frame of the unification model.

\end{abstract}

\section{Introduction}

Historic light curves of some bright BL Lac objects have shown that
fast luminosity fluctuations (typical of this class of AGNs) are
frequently superinposed on long-term trends of relatively large amplitude.
The origin of these long-term changes is not fully understood. An interesting
hypothesis is that they can be related to changes of the jet orientation
due, for instance, to the presence of a massive black hole binary system
in the nucleus of the object (Abraham and Carrara 1998; Lehto and Valtonen
1996).
We expect, therefore, that radio jets associated with these sources
should exibit large structural changes, detectable in VLBI images.
To this aim we started to observe some BL Lac objects, which are intensively
monitored in the optical, with the European VLBI Network (EVN).
One of them is ON~231 (W Com, 1219+285), discovered by
Wolf (1916) as a variable star and identified as the counterpart of a radio
source by Biraud (1971) and Browne (1971).
The historic light curve (Tosti et al. 1999a), although with gaps
before 1970, shows that the mean source luminosity
had a minimum in the early seventies and since that epoch has increased
to reach its highest level in the spring of 1998, when the source
also had a very bright outburst (Massaro et al. 1999).

The VLBI structure of ON~231 has been known for almost twenty years. 
The first 5 GHz image (observation epoch 1982.25) was presented by Weistrop
et al. (1985) and showed a core, emitting about one third of the total
flux, and a multicomponent jet elongated in Position Angle (PA)
$\sim$110$^\circ$.
The same structure was confirmed by subsequent observations
(Gabuzda et al. 1992, Gabuzda et al. 1994, Gabuzda \& Cawthorne 1996).
A recent image at 15\,GHz can be found in Kellermann et al. (1998) in
which the jet can be tracked up to about 10 mas from the nucleus.
The innermost structure at 22 GHz, with a resolution of about
0.35 mas, can be seen in the three VLBA maps covering the period from
1995.31 to 1997.58 presented by Jorstad et al. (2001).
Radio images of ON~231 on the arcsecond scale (de Bruyn \&
Schilizzi 1984, Kollgaard et al. 1992, Perlman \& Stocke 1994)
show emission from an extended faint region directed southwest (centred
at PA $\sim$310$^\circ$) with respect to the core.

We used the EVN$+$MERLIN facility to observe ON~231 at 1.6 GHz
and at 5 GHz. In this paper we present images made at the two 
frequencies
which show relevant new features in the source structure 
compared
to the previous published images. We propose some possible 
scenarios
to relate these changes with the long-term luminosity evolution.

\begin{figure}
\centerline{
\hbox{
\psfig{figure=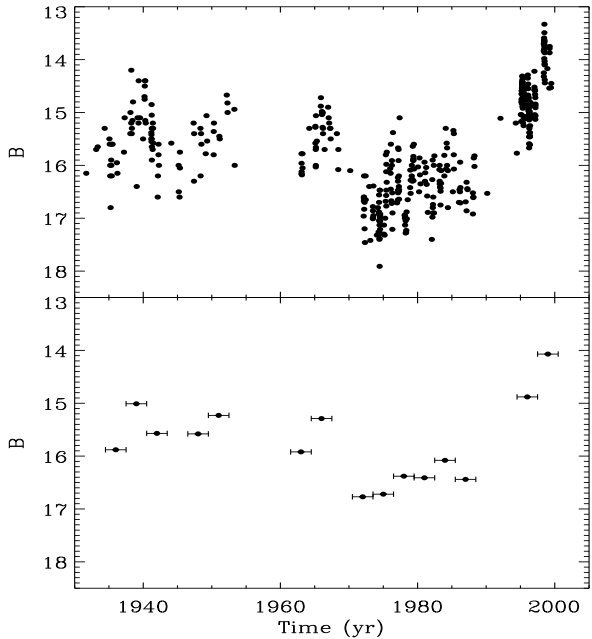,height=9cm,width=9cm,clip=}
}}
\caption{The historic light curve of ON~231 in the $B$ band after 1930.
The data up to the spring of 1997 are taken from Tosti et al. (1998),
while those of 1998 are given in Tosti et al. (1999b). In the lower panel
the same data have been averaged over intervals of three years.
}
\end{figure}

\section{Long-term behaviour in the optical and radio}

ON~231 is one of the few BL Lac objects whose variations have been known
for about one century, altough before the 1970s it was observed
only occasionally. For the past six years it has been intensively monitored,
in particular by the groups of Perugia and Rome.
Our photometric data up to 1998 have already been published
together with those of other collaborating groups in several papers
(Tosti et al. 1998, Tosti et al. 1999b, Massaro et al. 1999).
Additional recent data of the 1999 observational
season have been obtained with the Automatic Telescope of the
Perugia Observatory and the reflector of the Vallinfreda
Astronomical Station, near Rome. Observational equipment and
data reduction procedures are described in detail in the above
references. The up-to-date historic light curve is reported in
the upper panel of Fig.\,1.

The sparse observations of ON~231 before 1970 show
magnitude variations generally in the interval $14.5 - 16.5$. It
had a minimum around 1975 and after a few years a
brightening phase began. This brightening trend is clearer
if the data are averaged over time intervals longer than the
typical fluctuation duration. In Fig.\,1 (lower panel) we plotted
the mean magnitude computed over three year intervals: the
brightening trend after 1975 is evident despite the gap from
1990 to 1995.
The luminosity evolution in the period 1994--1999 is shown in detail
by the light curve in the R band of Fig. 2. Apart from the time
intervals in which ON~231 is not visible because of its proximity
to the sun, this light curve shows that the source behaviour
after 1995 was characterized by a series of major bursts, at least
one per year with the present time sampling, on which rapid
variations are superinposed. These events had a typical time-scale of
about five-six months and flux changes of a factor
of about 3. In Spring 1998, ON~231 had an exceptional outburst
and reached the highest recorded brightness level. 
In this episode the energy spectral index in the optical band 
hardened from --1.4 to --0.5 at the maximum and the linear polarisation 
increased up to about 10 \% with a PA in the range 
of 130$^\circ$--160$^\circ$ (Massaro et al. 1999), as expected in the case 
of synchrotron radiation from freshly accelerated electrons. 
After the burst, the mean source luminosity decreased and was 
comparable to that observed in the previous years, but with variations on 
shorter time scales.

\begin{figure}
\centerline{
\hbox{
\psfig{figure=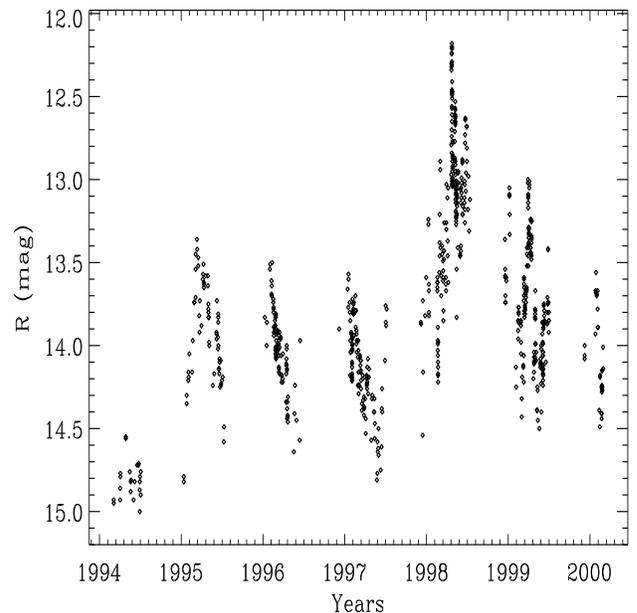,height=9cm,width=9cm,clip=}
}}
\caption{The light curve of ON~231 in the R (Cousins)
band after 1994.
}
\end{figure}

Radio light curves at various frequencies in the range from 4.8\,GHz
to 22\,GHz (Aller et al. 1999, Ter\"asranta et al. 1998) show
a behaviour not correlated with that in the optical. Fig.\,3 gives an
up-to-date version of the multiband light curve to the end of 1999
taken from the University of Michigan Radio Astronomy Observatory database.
ON~231 showed a general decreasing trend of the radio luminosity
starting from about 1980. After a four year long flare with a peak in
1992, the source flux started  to increase again in 1995.
The comparison with the optical light curve, despite the different
behaviour in the first portion, indicates that in 1995 ON~231 likely
entered a new phase in which both the optical and radio
flux levels started to increase.
Recall that our present knowledge about possible correlations
between the long-term behaviour of BL Lac objects in the radio and
optical bands is rather poor because of the small number of studied
sources and of the irregular time coverage. It is therefore hard to
reach a firm conclusion regarding whether the increase of the radio and optical
fluxes of ON~231 after 1995 is just a coincidence or it has a real
origin. The latter possibility can be adopted as a working
hypothesis, and can be used to formulate some scenarios for the evolution
of the jet geometry, as we will discuss in the last Section.

%on231_vlbi_f3.ps
\begin{figure}
\centerline{ \hbox{
\psfig{figure=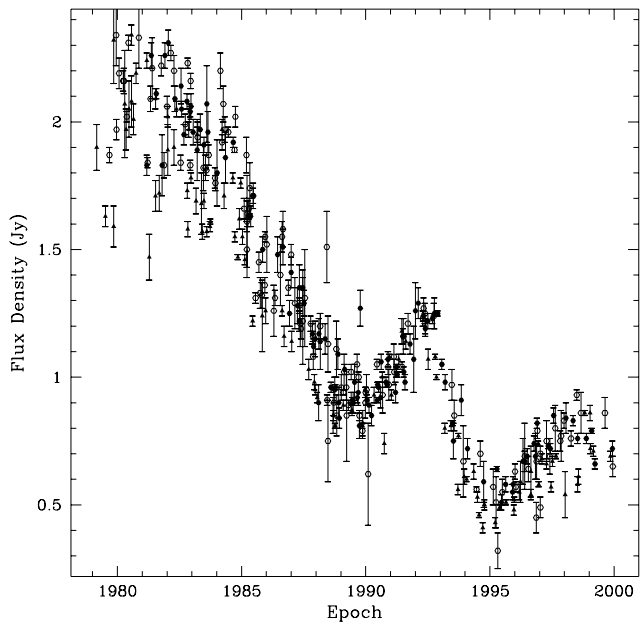,height=9cm,width=9cm,clip=} }}
 \caption{The radio light curve of
ON~231 from the UMRAO data base: 4.8\,GHz filled circles;8.0\,GHz
open circles; 14.5\,GHz triangles. }
\end{figure}

\section{ VLBI and MERLIN observations and data reduction}

ON\,231 was observed with the EVN plus
MERLIN on 05 June 1998 at 1.6\,GHz and on 17 February 1999 at 5\,GHz.
The data were recorded with the MarkIIIA terminals at the stations in
Mode B (28\,MHz bandwidth). During each experiment, the source was tracked
for about 11 hours together with the calibration sources OQ\,208 and 3C\,286,
observed for 13--minutes scans, regularly spaced over the experiment.
The VLBI data were correlated at the
MarkIIIA correlator of the Max-Planck-Institut f\"ur Radioastronomie
(Bonn, Germany). The raw data, output from the correlator, were integrated
for 4 seconds. Of the 11 stations of the EVN, namely Jodrell Bank and
Cambridge (UK), Westerbork (NL), Effelsberg (D), Medicina and Noto (I),
Onsala (S), Sheshan (Shanghai, China), Nanshan (Urumqi, China), Torun (PL),
Simeiz (Ukraine), nine gave fringes at 1.6\,GHz (all but Sheshan and Torun).
At 5\,GHz Effelsberg could not observe because of snow and Simeiz could not
take part in the observations. At the Jodrell Bank Observatory, the Lovell
telescope was used at 1.6\,GHz and the Mark2 telescope at 5\,GHz.

The correlator output was calibrated in amplitude and phase using
${\cal AIPS}$
\footnote {${\cal AIPS}$ is the NRAO's {\it Astronomical Image Processing
System}} and imaged using DIFMAP \footnote {DIFMAP is part of the
{\it Caltech VLBI software Package}} (Shepherd et al.\ 1995).
Total power measurements taken at the same times as the VLBI observations and
the gain curve of the telescopes were applied in the amplitude calibration
process.
The self-calibration procedure, which uses closure amplitudes to determine
telescope amplitude corrections, gave calibration factors that were within
10\% of unity for all the telescopes but Cambridge, Simeiz and Torun, which
were in the range 15-20\%.

The source was also tracked by MERLIN. Initial values for the telescope
and correlator gains were determined from a short observation of a
bright, unresolved calibration source. The primary calibrator 3C~286 was
scheduled during the observations in order to fix the absolute flux density
scale.
Images from the MERLIN LHC polarization data were produced using ${\cal AIPS}$.
An ${\cal AIPS}$ analysis path for processing joint EVN$+$MERLIN data was then
followed and the relative images were made using DIFMAP.

\section{ The structure of ON~231 on the parsec scale}

The complex structure of ON~231 is evident from the two full-resolution
VLBI images of  Figures 4 and 5 at 1.6\,GHz and
5\,GHz respectively. The source shows a straight collimated East-West
jet at PA $\sim$ 110$^\circ$ which bends twice by $\sim90^\circ$,
in a spiral-like structure. This extension South of the jet was not
present in previously published images, but it was partially detected
by observations with the VLBA at 2.3\,GHz in January 1997
\footnote{The VLBA images at 2.3\,GHz and
8.4\,GHz of 1219$+$285 are available in the Radio Reference Frame Image
Database (RRFID) of the United States Naval Observatory (USNO).
http://maia.usno.navy.mil/rorf/rrfid.html },
included in the Radio Reference Frame Image Database (RRFID).
Charlot, Sol and Vicente (1997) obtained a 5\,GHz image with
EVN observations in February 1994. The achieved mean beam size of about
20 mas was not adequate to resolve the jet but good enough to show the
presence of a component South-East of the core, which is likely the jet
extention seen in our images.

\begin{figure}
\centerline{
\hbox{
\psfig{figure=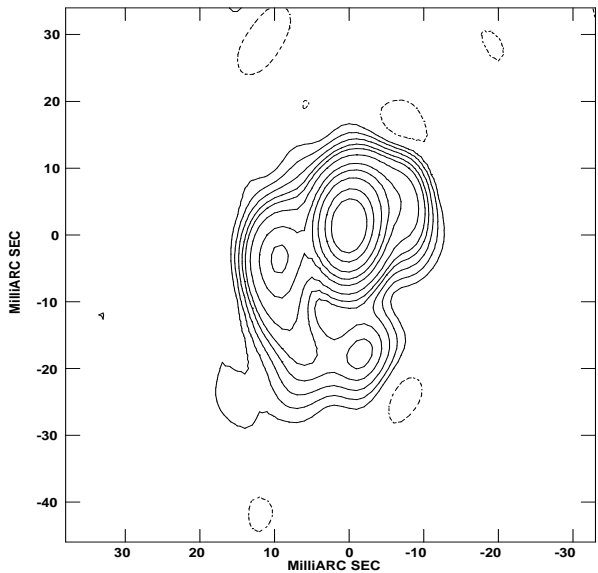,height=8.0cm,width=8.0cm,clip=}
}}
\caption{VLBI map of ON~231 at 1.6\,GHz. The contour levels are -3, 3, 
6, 10,
15, 20, 30, 50, 70, 100, 150, 200 mJy. The peak brightness 284.1
mJy beam$^{-1}$. The beam is 10.3$\times$4.8 mas$^2$ at P.A. 0$^\circ$.
}
\end{figure}

The jet shows a multicomponent structure and  
the most relevant difference with respect to literature
images (Gabuzda et al. 1992, 1994, Gabuzda \& Cawthorne 1996) 
is that the brightest component is no longer
at the western extreme of the source, but rather in a central location.
In fact, we clearly detect an extension on the western
side of the brightest component at both frequencies. Such an extension 
is also visible on a larger scale. The phase referenced MERLIN image at 
5 GHz shows that the source is resolved by the interferometer at a 50 mas 
resolution scale.
In Fig. 6 we can see an extension of the brightness distribution 
on both sides of the central bright component, along the source major axis 
at PA$\sim$110$^\circ$, plus an extension on the South-Eastern part of 
the source. 
We performed several tests to ensure that the western extension detected 
on the mas scale (component C0 in the 5 GHz EVN image) is real.
In the process of producing the hybrid radio images at both frequencies,
the self-calibration was done, windowing out the clean components
from the western emission.
The images we obtained were worse than those presented in the paper,
showing a poor fit between model and visibilities on the longest
baselines. In particular, the amplitude gain corrections required
for the two Chinese stations (which allowed high
resolution in the East-West direction) were of the order of $\sim$25\%
at 5 GHz.
In the 'normal' process, i.e. cleaning on the emission from the western
area persisting in the residual map, the corrections were $\leq 8\%$,
comparable to those applied to the other stations and in agreement with
the corrections required when self-calibrating the data set of the calibrator 
source OQ 208. We note that the images were made usinge both the analysis 
packages ${\cal AIPS}$ and DIFMAP, obtaining an
almost identical structure for 
ON~231.

These results raise the question of a proper identification of the
source core. Two hypotheses can be considered: $i)$ the
core is the brightest component and we observed the development
of a new component on the opposite side of the jet, $ii)$ the core
is located at the western extreme of the jet and we observed an
exceptionally bright intermediate component. To tackle the problem, 
we produced two further images by combining EVN and MERLIN 
data at both frequencies.
The images were made with the same resolution and were used to map the
distribution of the spectral index ($S \propto \nu^{\alpha}$),
reported in grey scale in Fig. 7, superposed on the
contour map at 5\,GHz. The uncertainties in the phase centre
locations in the two images (lost in processing the data), are at
the pixel level and can slightly affect the spectral index estimate.
Of course, we assumed that the mean flux of the various components
did not change dramatically in the eight months between the observations, 
as expected from the behaviour at these intermediate radio frequencies, 
shown in Fig.\,3.
The measured overall flux densities at both frequencies
in the MERLIN images and EVN$+$MERLIN images agree to better
than 1\%. We measured total fluxes of 903.3 mJy and 755.5 mJy at
1.6\,GHz and at 5\,GHz respectively, corresponding to an overall
spectral index of --0.16. The only component showing a flat
spectrum ($\alpha=0.06$) is the brightest one, labelled $b$ in
Fig. 7. We point out that the value for the East-Southern rim could 
be affected by the relatively low signal to noise ratio. 
The spectral index values of the other, more prominent components, 
labelled $a$, $c$ and $d$, are --0.60, --0.57 and --0.89, respectively.
This result suggests that the core is likely the brightest component
in $b$, i.e. component $C1$ in Fig. 5, rather than $C0$ at the western
edge of $b$, although we cannot rule out the latter hypothesis.

We performed a test to confirm the identification of $C1$ with the core.
Making use of the ${\cal AIPS}$ task JMFIT, we evaluated the angular 
distances between the main jet components in the low resolution 
EVN+MERLIN images. One would expect that if the core is actually located 
at the western edge of $b$, the angular distance from the bright, isolated, 
component $c$ in Fig.\,7 (corresponding to $C3$ in Fig\,5) and the core 
component $b$, which is a blend of $C0, C1$ and $C2$, should be greater 
in the image 
at 5\,GHz than in the image at 1.6\,Ghz. In fact, in the former case, 
$C0$, supposed to be the core, should contribute much more than $C2$ (almost
equal in flux density) to the total flux density of $b$ due to its flat or 
inverted spectral index between 1.6\,GHz and 5\,GHz, producing a
shift of the centre of the gaussian component(s) fitting $b$.
To check this we have tried to fit the brightness distribution of 
the main part of the jet region with a three-gaussian component model, 
leaving their central positions as free parameters. 
The position of the most Eastern component $c$ was then taken as 
reference point.  
In the 1.6\,GHz case we found that the centres of the two gaussians other 
than $c$, were located in PA$\sim 70^\circ$ at distances 
of 10.4 and 12.6 mas respectively. The solution for the brightness 
distribution at 5\,GHz failed to converge to the three gaussian components 
of the input model. The output was a two component model  
separated by 10.4 mas each other, equal to the separation found 
for a pair of components in the previous solution. The above solutions were 
found to be a stable convergence of the model fit process achieved by JMFIT 
in both cases, and suggest $C1$ as the core of ON\,231.
A further point in favour of this suggestion comes from the analysis of
the behaviour of the spectral indices of the individual components 
$C0, C1$ and 
$C2$, which give a total flux density of 454\,mJy at 5\,GHz for $b$ 
(from the model in Table \,1) and conspire to give a flat spectrum to it.
Let us take $C0$ as the core component in ON\,231 with an inverted spectral 
index of +2.5, as expected for a synchrotron self-absorbed source, and a 
spectral index $-0.7$ for $C2$. $C1$ will then contribute for about 260\,mJy 
to the total flux density of $b$ at 1.6\,GHz, implying it has an almost flat 
spectral index, rather unusual for a jet component. A more normal inverted 
spectral index for $C0$, say $\alpha=+1$, makes the spectral index of 
the two other components even flatter.
All these results indicate that the core is likely the brightest 
component $C1$ in $b$ rather than $a$ at the western edge, even though 
we cannot rule out the alternative hypothesis. 
We will further discuss this point in section 4.1.

The total jet extent along the axis in our images is $\simeq$ 15 mas,
while it is shorter in the previous images, ranging from about 9 to 11 mas.
A weak component at 14 mas is visible in the 8.4\,GHz image by Gabuzda
\& Cawthorne (1996) and a much farther one, at about 25 mas along the jet 
axis
at PA $\sim$ 110$^\circ$, was reported by Gabuzda et al. (1994).
We found no evidence for emission at such long distances from
the source centre.
We stress that according to the core identification given above, the
extension of the eastern jet would reduce to about 13\,mas, closer to the
previous estimates.

\begin{figure}
\centerline{
\hbox{
\psfig{figure=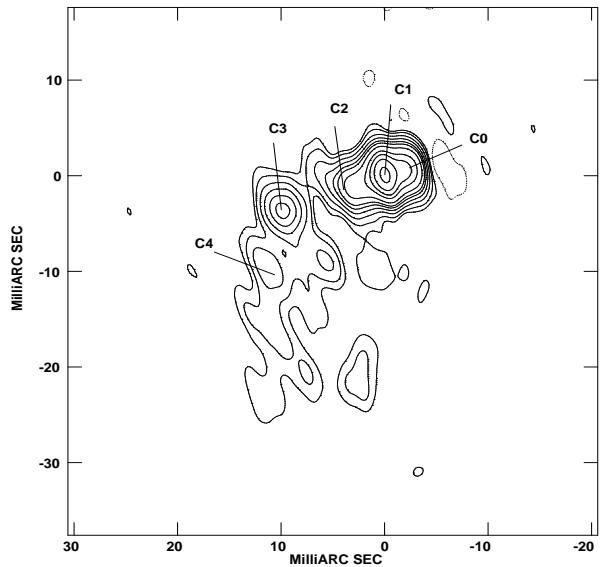,height=8.0cm,width=8.0cm,clip=}
}}
\caption{VLBI map of ON~231 at 5\,GHz. The contour levels are
-1.5, 1.5, 3, 6, 10, 15, 20, 30, 50, 70, 100, 150 mJy.
The peak brightness is 170.17 mJy beam$^{-1}$. The beam is
3.8$\times$1.4
mas$^2$ at PA 13$^\circ$.
}
\end{figure}

\begin{figure}
\centerline{
\hbox{
\psfig{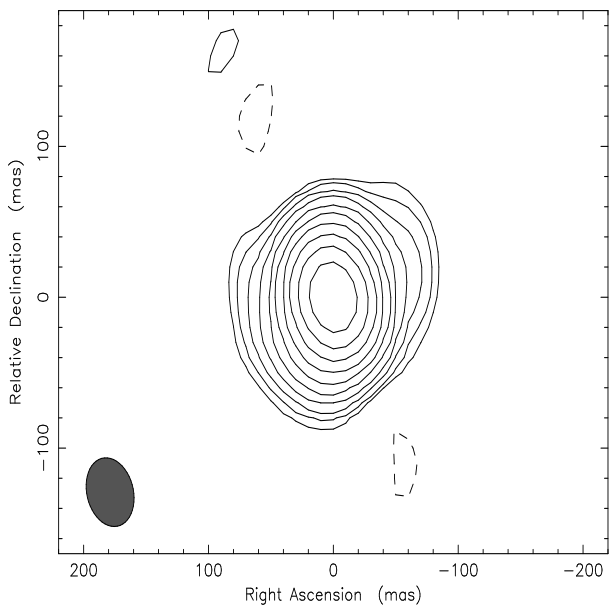}
}}
\caption{MERLIN image of ON~231 at 5\,GHz. The contour levels are -0.1
0.1, 0.2, 0.4, 0.8, 1.6, 3.2, 6.4, 12.8, 25.6, 51.2 \% of the
peak brightness of 672.6 mJy beam$^{-1}$. The beam is 47$\times$37 mas$^2$
at P.A. 23$^\circ$.
}
\end{figure}

\begin{figure}
\centerline{
\hbox{
\psfig{figure=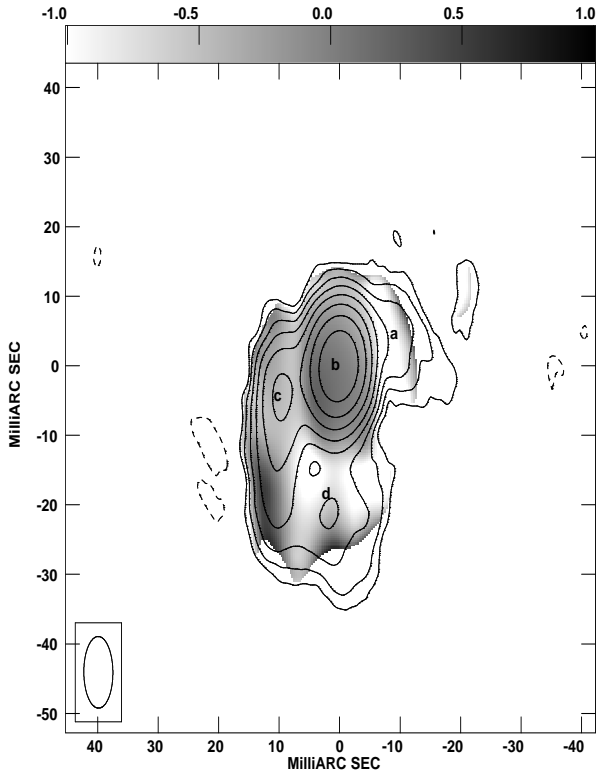,height=10.5cm,width=8.0cm,clip=
}
}}
\caption{Spectral index (grey scale) map computed from the EVN-MERLIN
combined images at 1.67 and 5 GHz, with a resulting beam of 10.3 $\times$
4.8 mas at PA 0$^\circ$. The 5 GHz contours correspond to
-0.4, 0.4, 0.8, 1.6, 3.2, 6.4, 12.8, 25.6, 51.2, per cent of the peak
brightness of 426.4 mJy beam$^{-1}$.
}
\end{figure}

\subsection{ Components of the jet structure }

The occurrence of various emitting components in the jet of ON~231 has
been recognized from early VLBI images.
Weistrop et al. (1985) reported the detection of at least three components.
Gabuzda et al. (1992) detected four components, later increased to
eight/nine by Gabuzda et al. (1994) and Gabuzda \& Cawthorne (1996).
From these observations, which span a time interval of about three years, 
the authors estimated component velocities $\beta_{app} h$ in the range 
0.62 -- 2.42, increasing with the distance from the core.
Jorstad et al. (2001) at 22\,GHz found a component moving at
1.5 -- 2.3 in the inner part of the jet.

We present in Fig.\,5 the most resolved source structure at 5\,GHz  that
can be modelled with five gaussian components, with a good formal agreement 
between data and model,with $\chi^2_{\nu}$ quite close to unity. In Table\,1, 
together with the flux of each component, we report the angular distances 
from the centre of the brightest component $C1$, the corresponding PAs, 
the major axis, the axial ratios and the orientation of the best fitted gaussians.
The component $C1$ is the one giving the major contribution to the flux
of the flat spectrum component $b$ in the map of Fig.\,7. Taking into account
that both components are in the same position despite the different resolution,
we can identify $C1$ as the core of ON~231. 
The model parameters of Table \,1 fit well the visibilities of the longest
baselines at 1.6\,GHz, while it does not account for the total flux density on
the short baselines.

\begin{table}[h]
\centerline{\bf Table\,1 - Modelfitting 5\,GHz data set}
\vspace{0.5cm}
\hspace{0.5cm} %if you want to center your table act on this argument
\begin{tabular}{crcrccr}
Comp.         & Flux & Radius & PA     & Major  & Axial & Phi        \\
              & (mJy)& (mas)  & (deg)  & (mas)  & ratio & (deg)      \\
\hline
{\it C}0       & 110.6&  2.36  & $-$71.6& 1.8    & 0.32  & $-$86.1    \\
{\it C}1       & 243.0&  0.00  & 0.0    & 1.5    & 0.45  & $-$81.1    \\
{\it C}2       & 100.3&  3.32  & 109.0  & 3.0    & 0.67  & 88.6       \\
{\it C}3       &  66.6& 10.4   & 113.4  & 5.7    & 0.45  & 21.8       \\
{\it C}4       &  19.5& 14.7   & 135.8  & 3.7    & 0.60  & $-$76.2    \\
\hline
\end{tabular}
\vspace{0.5cm}
$Note$: components are ordered from West to East
\end{table}

\subsection{ Jet orientation and intrinsic speed }

In the following, we derived some estimates for the source orientation with
respect to the line of sight and for the intrinsic plasma speed based on a 
comparison of our images and previously published data used to derive proper 
motion estimates. We also investigate the possibility that the steep-spectrum
emission to the west of $C1$ is indicative of the counter-jet. For our purposes
we used the full resolution 5\,GHz image reported in Fig.\,5 and the components
from the model fitting given in Table\,1.

It is not easy to support a proper association between any of the components
in our model and those from observations made roughly 10 years ago
by Gabuzda et al. (1994). The brightness profile along the jet in our images
is quite different from those resulting from previous observations,
as shown in Fig.\,8. Note that the peak flux of the component $C1$, which
we identify as the core of the radio emission, decreased from 250 mJy in the 
early observations by Gabuzda et al. (1992)
to 170 mJy in our data with a 
ratio in agreement with the total flux decrease in this time interval (Fig.\,3).

A plausible association could be made between our component $C3$
and the $K2$ of Gabuzda et al. (1994). According to their study, this component
moved eastward at $\sim$ 0.37 mas/year in 1.88 years. Assuming that its velocity 
remained
constant in the following ten years, it should now be at about 11.3 mas from
the core, compatible with the distance of $C3$ from the component $C1$.

A proper motion $\mu =$ 0.37 mas/year corresponds to $\beta_{app}=$ 1.66 for H$_o$ =
100 km s$^{-1}$ Mpc$^{-1}$ and q$_o$=0.5 (same cosmology as in Gabuzda et al. 1994).
using the standard formulae given in Pearson \& Zensus (1987), the derived apparent
superluminal speed implies a minimum value for the intrinsic plasma speed 
$\beta_{intr}=$ 0.84,
which can be obtained if the viewing angle $\theta \sim 31^\circ$. The maximum allowed
angle is $\sim$ 62$^\circ$, which is a large value for BL Lac, even considering that it is
an upper limit.

Let us now test whether it is plausible that the steep-spectrum emission 
to the west of
$C1$ is a counter-jet; i.e. the Doppler de-boosted jet that is receding 
from the observer.
If we assume that the intrinsic source structure is symmetrical, an estimate 
of $\beta_{intr}$
and $\theta$ can be derived. If $C1$ is the core, in order to estimate the 
jet to 
counter-jet brightness ratio $R(S)$, we should compare the flux of the 
component $C0$
to the sum of the fluxes from $C2$, $C3$ and $C4$. This leads to a ratio 
$R(S) \sim 1.7$,
corresponding to $\beta$ cos$\theta$ $\sim$ 0.11. As indicated above, 
based on the observed
apparent superluminal motion, $\beta_{intr} \geq $ 0.84; this implies a 
$maximum$ value of
$cos \theta \sim$ 0.13, or a minimum viewing angle of 
$\theta \sim$ 82$^\circ$. This value
is far too large to be plausible; in other words, the emission to the 
West of $C1$ is far
too bright to be a counter-jet. We discuss alternative interpretations of 
this emission
below.

\begin{figure}
\centerline{
\hbox{
\psfig{figure=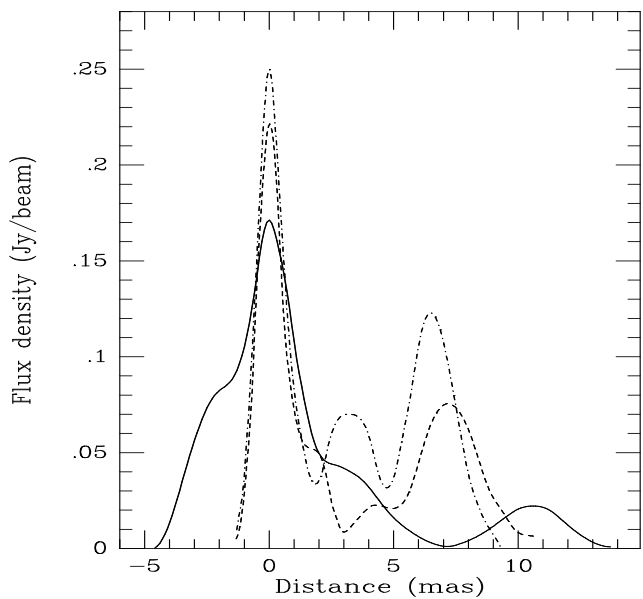,height=9.0cm,width=9.0cm,clip=}
}}
\caption{5\,GHz intensity profiles along the jet at PA $\sim$ 110$^\circ$
at different epochs: 1987.41 (dot-dashed line) from Gabuzda et al.(1992),
1989.29 (dashed line) from Gabuzda et al. (1994) and 1999.13 (solid line),
our data.
}
\end{figure}

\section{ Discussion }

The intensive optical monitoring of some bright BL Lac objects shows
the existence of long-term trends, which can be related to structure
modifications of the source.
Our recent VLBI images of ON~231 were obtained during the very bright and
active phase in the optical beginning in 1995, following the twenty--year 
increase in the mean luminosity, and show significant changes with respect
to several previous VLBI images taken since 1987.

As already discussed in Sect. 4 we identified the source core
with the brightest component on the basis of its flat
spectrum. The most important consequence of this result is
that we detected a new component, not visible in older images, 
at PA $\sim$--70$^\circ$, opposite to the main jet.
The absence of this component in the subsequent 22 GHz map (epoch
1997.58) by Jorstad et al. (2001) could be due to its
rather steep spectrum. We stress, however, that those images
indicate a brightening of a component at about 2 mas East of the 
core whose flux density remained the strongest.

Parsec-scale jets of BL Lac objects are generally one-sided, and 
this is explained by the strong Doppler boosting of the jet pointing 
to the observer line of sight and by the corresponding deboosting of 
the jet at the opposite direction.
In a few cases, however, indication for double jet and complex structures
have been reported in the literature, like those found in 1308+326 
by Pyatumina, Aller \& Aller (1999), who analysed a series of 8 GHz 
images from the geodetic VLBI data base. In particular, they noticed the 
appearance of components in different directions.
Furthermore, jet structures in BL Lac sources are frequently not very
well aligned and show wiggles, suggesting either a strong interaction
with the surrounding medium or some kind of motion, regular (e.g.
precession) or not, of the jet itself.

We can consider a few possible scenarios for interpreting both the time
evolution of the optical luminosity and the changes of the jet structure
of ON 231 on the parsec scale. A first possibility is that the jet,
pointing very close to the observing direction, undergoes strong instabilities 
and oscillations, and therefore the new western
component is the result of one episodic large amplitude displacement. Again,
emitting blobs from the nuclear source can be distributed around a mean
direction of the motion and one of them could be displaced toward West at an 
angle greater than that of other blobs.
Future VLBI images will then show the evolution of this unique component, while 
the others will continue to move along the former direction. 
It will be interesting to verify if the occurrence of new very bright optical 
outbursts are related to the presence of components on the west side (or other directions) 
of the core.

Another possible scenario is that of a slowly precessing jet
approaching the observer's line of sight over the past few decades.
The progressive increase of the beaming factor would then be
responsible for the mean brightening optical trend shown by this source,and 
the minimum angular distance was likely
reached in 1997-98 when the 
source was observed at the highest level. 
After this phase, the apparent jet direction changed to
the opposite side as in our images. 
A more detailed model, which takes into account the evolution of the
emitting electron population and the radiation transfer, is necessary to
explain the absence of a positive correlation between the radio and
optical light curves, but it is beyond the aim of the present paper.
In this framework, we can expect that in the next five or ten years a further 
development of the western jet
and a fading of the eastern jet will be oserved.
The absence of radio emission on the North side of the jet axis,
even on the arsecond scale (Kollgaard et al. 1992, Perlman \& Stocke 1994),
indicates that the precession cone axis lies South-Southwest
respect to the line of sight.

If the above core identification is not correct and the core corresponds
to the weak component at the western extreme, it is necessary to understand 
why a jet component achieved a spectrum flatter than the core itself.
A possible explanation is that the jet suffered a strong interaction with
the surrounding matter and changed its direction, as in the recent model
proposed by Pohl \& Schlickaiser (2000). A relativistic hadronic blast wave
is emitted by the central engine and a radiation burst is produced when
it interacts with near dense clouds. Synchrotron emitting electrons can
be originated from the decay of charged pions produced in the nuclear
collisions between the blast wave particles and those of the cloud. We expect,
therefore, that gamma rays having energies greater than about 50 MeV are
copiously produced in such an event. ON~231 is known to have been detected by
EGRET (Mukherjee et al. 1997), but unfortunately no data after 1995.5
are available.

The extension of the jet after the bend is clearly detected in our images
for the first time.
Since this is not a new emitting region, it was not detected in previous
images due to sensitivity limits. In the above scenario of a precessing jet,
it could be the trace of the jet itself, rotating clockwise from South to
North. A second possibility is that the bending of the jet is the consequence
of a strong interaction with the near environment.

Finally, we would like to stress the relevance of a coordinated study
of multifrequency variability with VLBI mapping, particularly for BL 
Lac objects like ON~231 which have the synchrotron peak
in the near IR-optical band. Our data suggest that the
occurrence of a long-term trend in the optical luminosity and of periods
of enhanced activity could be related to changes in the innermost radio 
structure.
A better understanding of these phenomenon requires both an intense optical
monitoring and VLBI mapping, spaced not longer than one year apart. Such a program
can be realised only for a small sample of sources and ON~231 should be
one of the primary targets.

\begin{acknowledgements}
The Roma and Perugia groups acknowledge the financial support by the
Italian Ministry for University and Research (MURST) under the grant
Cofin 98-02-32.
We acknowledge support from European Commission's TMR-LSF
programme, Contract No. ERBFMGECT950012.
This research has made use of the data from the University of Michigan
Radio Astronomy Observatory which is supported by funds  from the
University of Michigan.
This research has made use of the United States Naval Observatory (USNO)
Radio Reference Frame Image Database (RRFID).

\end{acknowledgements}

\end{document}